\documentclass[12pt,twoside,a4paper]{article}
\usepackage{amsmath, amsthm, amssymb}

\usepackage{rotating}
\usepackage[dvips]{epsfig}
\newcommand{\gsim}{\,\rlap{\lower 3.5 pt \hbox{$\mathchar \sim$}} \raise 1pt
 \hbox {$>$}\,}
\newcommand{\lsim}{\,\rlap{\lower 3.5 pt \hbox{$\mathchar \sim$}} \raise 1pt
 \hbox {$<$}\,}

\catcode`@=11
\newcount\@tempcntc
\def\@citex[#1]#2{\if@filesw\immediate\write\@auxout{\string\citation{#2}}\fi
  \@tempcnta\z@\@tempcntb\m@ne\def\@citea{}\@cite{\@for\@citeb:=#2\do
    {\@ifundefined
       {b@\@citeb}{\@citeo\@tempcntb\m@ne\@citea\def\@citea{,}{\bf ?}\@warning
       {Citation `\@citeb' on page \thepage \space undefined}}%
    {\setbox\z@\hbox{\global\@tempcntc0\csname b@\@citeb\endcsname\relax}%
     \ifnum\@tempcntc=\z@ \@citeo\@tempcntb\m@ne
       \@citea\def\@citea{,}\hbox{\csname b@\@citeb\endcsname}%
     \else
      \advance\@tempcntb\@ne
      \ifnum\@tempcntb=\@tempcntc
      \else\advance\@tempcntb\m@ne\@citeo
      \@tempcnta\@tempcntc\@tempcntb\@tempcntc\fi\fi}}\@citeo}{#1}}
\def\@citeo{\ifnum\@tempcnta>\@tempcntb\else\@citea\def\@citea{,}%
  \ifnum\@tempcnta=\@tempcntb\the\@tempcnta\else
   {\advance\@tempcnta\@ne\ifnum\@tempcnta=\@tempcntb \else \def\@citea{--}\fi
    \advance\@tempcnta\m@ne\the\@tempcnta\@citea\the\@tempcntb}\fi\fi}
\catcode`@=12

\begin{document}
\thispagestyle{empty} 
\title{
\vskip-3cm
{\baselineskip14pt
\centerline{\normalsize DESY~09--086 \hfill ISSN 0418--9833}
\centerline{\normalsize June 2009 \hfill}} 
\vskip1.5cm
\boldmath
{\bf Small-$x$ behavior of the structure function $F_2$ and its slope 
$\partial\ln F_2/\partial\ln(1/x)$ for ``frozen'' and analytic strong-coupling
constants}
\unboldmath}
\author{G.~Cveti\v{c}$^1$, A.~Yu.~Illarionov$^2$, B.~A.~Kniehl$^3$,
A.~V.~Kotikov$^3$\thanks{%
On leave of absence from Bogoliubov Laboratory of Theoretical
Physics, Joint Institute for Nuclear Research, 141980 Dubna, Moscow region,
Russia.}
\bigskip\\
{\normalsize $^1$ Department of Physics, Universidad T\'ecnica Federico Santa
Mar\'{\i}a,}
\\
\normalsize{Avenida Espa\~na 1680, Casilla 110--V, Valpara\'{\i}o, Chile}
\bigskip\\
{\normalsize $^2$ International School for Advanced Studies SISSA,}
\\
\normalsize{via Beirut 2--4, 34014 Trieste, Italy}
\bigskip\\
{\normalsize $^3$ II. Institut f\"ur Theoretische Physik, Universit\"at
Hamburg,}\\
\normalsize{Luruper Chaussee 149, 22761 Hamburg, Germany}
}

\date{}
\maketitle
\begin{abstract}
Using the leading-twist approximation of the Wilson operator product expansion
with ``frozen'' and analytic versions of the strong-coupling constant, we show
that the Bessel-inspired behavior of the structure function $F_2$ and its
slope\break $\partial\ln F_2/\partial\ln(1/x)$ at small values of $x$, obtained for
a flat initial condition in the DGLAP evolution equations, leads to good
agreement with experimental data of deep-inelastic scattering at DESY HERA.
\medskip

\end{abstract}

\clearpage

\section{Introduction} 
\label{sec:intro}

The experimental data from DESY HERA on the structure function $F_2$ of
deep-inelastic scattering (DIS) 
\cite{Abt:1993,Ahmed:1995,Aid:1996,Adloff:1997,Adloff:1999,Adloff:2001,%
Derrick:1993,Derrick:1995,Derrick:1996:C69,Derrick:1996:C72,Breitweg:1997,%
Breitweg:1999,Breitweg:2000,Chekanov:2001}
and its derivatives $\partial F_2/\partial\ln Q^2$
\cite{Adloff:1997,Adloff:2001,Surrow:2002}
and $\partial\ln F_2/\partial\ln(1/x)$
\cite{Surrow:2002,Adloff:2001rw,Lastovicka:2002,Gayler:2002}
bring us into a very interesting kinematic range for testing theoretical ideas
on the behavior of quarks and gluons carrying a very small fraction of the
proton's momentum, the so-called small-$x$ region.
In this limit, one expects that the conventional treatment based on the
Dokshitzer-Gribov-Lipatov-Altarelli-Parisi (DGLAP) equations
\cite{Gribov:1972:1000+,Lipatov:1975,Altarelli:1977,Dokshitzer:1977}
does not account for contributions to the cross section which are
leading in $\alpha_s\ln(1/x)$; moreover, the parton density functions (PDFs),
in particular the one of the gluon, become large, and the need arises to apply
a high-density formulation of QCD.

 However, reasonable agreement between HERA data and the next-to-leading-order
(NLO) approximation of perturbative QCD has been observed for
$Q^2\gsim 2$~GeV$^2$ (see reviews in
Ref.~\cite{Cooper-Sarkar:1998,Kotikov:2007ua} and references cited therein)
indicating that perturbative QCD can describe the evolution of $F_2$ and its
derivatives down to very small $Q^2$ values, traditionally characterized by
soft processes.

The standard program to study the $x$ dependence of quark and gluon PDFs is to
compare the numerical solutions of the DGLAP equations with the data and so to
fit the parameters of the $x$ profiles of the PDFs at some initial
factorization scale $Q_0^2$ and the asymptotic scale parameter $\Lambda$.
However, for analyzing exclusively the small-$x$ region, there is the
alternative of doing a simpler analysis by using some of the existing
analytical solutions of the DGLAP equations in the small-$x$ limit
\cite{Ball:1994,Mankiewicz:1997,Kotikov:1999,Illarionov:2008}.
This was done in Ref.~\cite{Ball:1994}, where it was pointed out that the
small-$x$ data from HERA can be interpreted in terms of the so-called
double-asymptotic-scaling (DAS) phenomenon related to the asymptotic behavior
of the DGLAP evolution discovered in Ref.~\cite{DeRujula:1974} many years ago.

The study of Ref.~\cite{Ball:1994} was extended in
Refs.~\cite{Mankiewicz:1997,Kotikov:1999,Illarionov:2008} to include the
subasymptotic part of the $Q^2$ evolution.
This led to predictions \cite{Kotikov:1999,Illarionov:2008} of the small-$x$
asymptotic PDF forms in the framework of DGLAP dynamics starting at some
initial value $Q^2_0$ with flat $x$ distributions:
\begin{equation}
	x f_a (x,Q^2_0) = A_a\qquad
(a=q,g),
\label{1}
\end{equation}
where $f_a(x,Q^2)$ are the PDFs and $A_a$ are unknown constants to be
determined from the data.
We refer to the approach of
Refs.~\cite{Mankiewicz:1997,Kotikov:1999,Illarionov:2008} as
\emph{generalized} DAS approximation.
In this approach, the flat initial conditions in Eq.~(\ref{1}) play the basic
role of the singular parts of the anomalous dimensions by determining the
small-$x$ asymptotics, as in the standard DAS case, while the contributions
from the finite parts of the anomalous dimensions and from the Wilson
coefficients can be considered as subasymptotic corrections, which are,
however, important for better agreement with the experimental data.
In the present paper, similarly to
Refs.~\cite{Ball:1994,Mankiewicz:1997,Kotikov:1999,Illarionov:2008}, we neglect
the contribution from the non-singlet quark component.

The use of the flat initial condition given in Eq.~(\ref{1}) is supported by
the actual experimental situation: small-$Q^2$ data
\cite{Adloff:1997,Adloff:2001,Breitweg:1997,Surrow:2002,Arneodo:1995,%
Arneodo:1997,Adams:1996gu}
are well described for $Q^2\leq0.4$~GeV$^2$ by Regge theory with Pomeron
intercept $\alpha_P(0)=1+\lambda_P=1.08$ (see Ref.~\cite{Donnachie:1998} and
references cited therein), close to the standard one, $\alpha_P(0)=1$. 
The small rise of the HERA data
\cite{Adloff:1997,Adloff:2001,Breitweg:1997,Breitweg:2000,Surrow:2002}
at small values of $Q^2$ can be explained, for instance, by contributions of
higher-twist operators \cite{Illarionov:2008}.

The purpose of this Letter is to compare the predictions for the structure
function $F_2(x,Q^2)$ and its slope\break $\partial\ln F_2/\partial\ln(1/x)$
from the generalized DAS approach with H1 and ZEUS experimental data
\cite{Abt:1993,Ahmed:1995,Aid:1996,Adloff:1997,Adloff:1999,Adloff:2001,%
Derrick:1993,Derrick:1995,Derrick:1996:C69,Derrick:1996:C72,Breitweg:1997,%
Breitweg:1999,Breitweg:2000,Chekanov:2001,%
Surrow:2002,Adloff:2001rw,Lastovicka:2002,Gayler:2002}.
Detailed inspection  of the H1 data points
\cite{Adloff:1997,Adloff:2001,Adloff:2001rw} reveals that, in the ranges
$x<0.01$ and $Q^2\gsim2$ GeV$^2$, they exhibit a power-like behaviour of the
form
\begin{equation}
	F_2(x,Q^2) = C x^{-\lambda (Q^2)},
\label{1dd}
\end{equation}
where the slope $\lambda(Q^2)$ is, to good approximation, independent of $x$
and scales logarithmically with $Q^2$, as $\lambda(Q^2)=a\ln(Q^2/\Lambda^2)$.
A fit yields $C \approx 0.18$, $a \approx 0.048$, and $ \Lambda = 292$~MeV
\cite{Adloff:2001rw}.
The linear rise of $\lambda(Q^2)$ with $\ln Q^2$ is also indicated in
Figs.~\ref{fig:Q2-slope} and \ref{fig:x-slope}, to be discussed below.

The rise of $\lambda(Q^2)$ linearly with $\ln Q^2$ can be traced to strong 
nonperturbative physics (see Ref.~\cite{Schrempp:2005} and references cited
therein), i.e.\ $\lambda (Q^2) \sim 1/\alpha_s(Q^2)$. 
However, the analysis of Ref.~\cite{Kotikov:1997:JETP} demonstrated that this
rise can be explained naturally in the framework of perturbative QCD (see also
Section~\ref{sec:results}).

The H1 and ZEUS Collaborations \cite{Surrow:2002,Lastovicka:2002,Gayler:2002}
also presented new data for $\lambda (Q^2)$ at quite small values of $Q^2$.
As may be seen from Fig.~8 of Ref.~\cite{Surrow:2002}, the ZEUS value for 
$\lambda(Q^2)$ is consistent with a constant of about 0.1 at
$Q^2\lsim0.6$~GeV$^2$, as is expected under the assumption of
single-soft-Pomeron exchange within the framework of Regge phenomenology. 

It is interesting to extend the analysis of Ref.~\cite{Kotikov:1997:JETP} to
the small-$Q^2$ range with the help of the well-known infrared modifications of
the strong-coupling constant. 
We shall adopt the ``frozen'' \cite{Badelek:1996} and analytic
\cite{Shirkov:1997} versions.

This paper is organized as follows.
Section~\ref{sec:DAS} 
contains basic formulae for
the structure function $F_2$ and its slope $\partial\ln F_2/\partial\ln (1/x)$
in the generalized DAS approximation
\cite{Kotikov:1999,Illarionov:2008,Kotikov:1997:JETP},
which are needed for the present study.
In Section~\ref{sec:results}, we compare our results on $F_2$ and
$\partial\ln F_2/\partial\ln(1/x)$ with experimental data.
Our conclusions may be found in Section~\ref{sec:concl}.

\section{Generalized DAS approach} 
\label{sec:DAS}

The flat initial conditions in Eq.~(\ref{1}) correspond to the case when the
PDFs tend to constants as $x \to 0$ at some initial value $Q^2_0$.
The main ingredients of the results at the leading order (LO)
\cite{Kotikov:1999,Illarionov:2008} include the following.\footnote{
The NLO results may be found in Refs.~\cite{Kotikov:1999,Illarionov:2008}.}
Both, the gluon and quark-singlet PDFs are presented in terms of two
components (``$+$" and ``$-$"), 
\begin{equation}
	F_2(x,Q^2)=exf_q(x,Q^2),\qquad
	f_a(x,Q^2)=f_a^{+}(x,Q^2) + f_a^{-}(x,Q^2)
\qquad (a=q,g),
\label{intro:1}
\end{equation}
which are obtained from the analytic $Q^2$-dependent expressions of the
corresponding (``$+$" and ``$-$") PDF moments.
Here, $e=(\sum_{i=1}^f e_i^2)/f$ is the average charge square and $f$ is the
number of active quark flavors.
The small-$x$ asymptotic results for the PDFs $f^{\pm}_a$ are
\begin{align}
	xf^{+}_q(x,Q^2) &= \dfrac{f}{9}\left(A_g + \dfrac{4}{9} A_q \right)
		\rho \tilde I_1(\sigma)  e^{-\overline d_{+}(1) s} + O(\rho),
\qquad
	f^{+}_g(x,Q^2) = \frac{9\tilde I_0(\sigma)}{f\rho\tilde I_1(\sigma)}
	f^{+}_q(x,Q^2),
	\nonumber \\
	xf^{-}_q(x,Q^2) &= A_q e^{-d_{-}(1) s} + O(x) ,
\qquad
	f^{-}_g(x,Q^2) = -\dfrac{4}{9}f^{-}_q(x,Q^2),
	\label{8.02}
\end{align}
where $\overline d_{+}(1) = 1 + 20f/(27\beta_0)$ and
$d_{-}(1) = 16f/(27\beta_0)$
are the regular parts of the anomalous dimensions $d_{+}(n)$ and $d_{-}(n)$, 
respectively, in the limit $n\to1$.\footnote{%
We denote the singular and regular parts of a given quantity $k(n)$ in the
limit $n\to1$ by $\hat k(n)$ and $\overline k(n)$, respectively.}
Here, $n$ is the variable in Mellin space.
The functions $\tilde I_{\nu}$ ($\nu=0,1$) are related to the modified Bessel
function $I_{\nu}$ and the Bessel function $J_{\nu}$ by
\begin{equation}
\tilde I_{\nu}(\sigma) =
\left\{
\begin{array}{ll}
I_{\nu}(\sigma) , & \mbox{ if } s \geq 0; \\
i^{-\nu} J_{\nu}(i\sigma) , & \mbox{ if } s < 0. 
\end{array}
\right.
\label{4}
\end{equation}
The variables $s$, $\sigma$, and $\rho$ are
given by
\begin{equation}
	s = \ln\dfrac{\alpha^{\rm LO}_s(Q^2_0)}{\alpha^{\rm LO}_s(Q^2)},\qquad
	\sigma = 2\sqrt{\hat{d}_{+}(s-i\epsilon) \ln x},\qquad
	\rho = \dfrac{\sigma}{2\ln(1/x)}, 
\label{slo}
\end{equation}
where $\hat{d}_{+}=-12/\beta_0$, $\alpha^{\rm LO}_s(Q^2)$ is the
strong-coupling constant in the LO approximation, and $\beta_0$ is the first
term of its $\beta$ function.

Contrary to the approach of
Refs.~\cite{Ball:1994,Mankiewicz:1997,Kotikov:1999,Illarionov:2008},
various groups were able to fit the available data using a hard input at small
values of $x$, of the form $x^{-\lambda}$, with different values $\lambda>0$
at small and large values of $Q^2$ 
\cite{Donnachie:1998,Abramowicz:1991,Kotikov:1992:YF,Jenkovszky:1993PL,%
Capella:1994,Frichter:1995,Kotikov:1996:MPL,Kotikov:1996:YF,Lopez:1996,%
Adel:1997,Kaidalov:2000,Donnachie:2003cs}.
At small $Q^2$ values, there are well-known such results
\cite{Donnachie:1998}.
At large $Q^2$ values, this is not very surprising for the modern HERA data
because they cannot distinguish between the behavior based on a steep PDF
input at quite large $Q^2$ values and the steep form acquired after the
dynamical evolution from a flat initial condition at quite small $Q^2$ values.
 
As has been shown in Refs.~\cite{Kotikov:1999,Illarionov:2008}, the $x$
dependencies of $F_2$  and the PDFs given by the Bessel-like forms in the
generalized DAS approach can mimic power-law shapes over a limited region of
$x$ and $Q^2$ values:
\begin{equation}
	F_2(x,Q^2) \sim x^{-\lambda^{\rm eff}_{F_2}(x,Q^2)},\qquad
 	x f_a(x,Q^2) \sim x^{-\lambda^{\rm eff}_a(x,Q^2)}.
\label{10.a}
\end{equation}
In the twist-two LO approximation, the effective slopes have the following
forms: 
\begin{equation}
	\lambda^{\rm eff}_{F_2}(x,Q^2) 
= \lambda^{\rm eff}_q(x,Q^2) = \dfrac{f^+_q(x,Q^2)}{f_q(x,Q^2)}
		\rho  \dfrac{\tilde I_2(\sigma)}{\tilde I_1(\sigma)},\qquad
	\lambda^{\rm eff}_g(x,Q^2) 
= \dfrac{f^+_g(x,Q^2)}{f_g(x,Q^2)} 
		\rho \frac{\tilde I_1(\sigma)}{\tilde I_0(\sigma)}.
\label{10.1}
\end{equation}
The corresponding NLO expressions and the higher-twist terms may be found in
Refs.~\cite{Kotikov:1999,Illarionov:2008}.

The effective slopes $\lambda^{\rm eff}_{F_2}$ and $\lambda^{\rm eff}_a$
depend on the magnitudes $A_a$ of the initial PDFs and also on the chosen
input values of $Q^2_0$ and $\Lambda$.
To compare with the experimental data, it is necessary to use the exact
expressions from Eq.~(\ref{10.1}), but for a qualitative analysis one can
use some appropriate approximations.
At large values of $Q^2$, the ``$-$'' components of the PDFs are negligible,
and the dependencies of the slopes on the PDFs disappear.
In this case, the asymptotic behaviors of the slopes are given by the
following expressions:\footnote{%
The asymptotic formulae given in Eq.~(\ref{11.1}) work quite well at any
values $Q^2 \geq Q^2_0$, because at $Q^2=Q^2_0$ the values of
$\lambda^{\rm eff}_a$ and $\lambda^{\rm eff}_{F_2}$ are equal to zero. 
The use of the approximations in Eq.~(\ref{11.1}) instead of the exact results 
given in Eq.~(\ref{10.1}) underestimates (overestimates) the gluon (quark)
slope at $Q^2 \geq Q^2_0$ only slightly.}
\begin{equation}
\lambda^{\rm eff,as}_q(x,Q^2)\approx \rho -\dfrac{3}{4\ln{(1/x)}},\qquad 
\lambda^{\rm eff,as}_g(x,Q^2)\approx \rho -\dfrac{1}{4\ln{(1/x)}}.
\label{11.1}
\end{equation}
where the symbol $\approx$ marks the approximation obtained from the expansion
of the usual and modified Bessel functions in Eq.~(\ref{4}).
One can see from Eq.~(\ref{11.1}) that the gluon effective slope
$\lambda^{\rm eff,as}_g$ is larger than the quark one
$\lambda^{\rm eff,as}_q$, which is in excellent agreement with global analyses 
\cite{Cooper-Sarkar:1998,Kotikov:2007ua}. 

\section{Comparison with experimental data} 
\label{sec:results}

\begin{figure}[t]
\includegraphics[width=\textwidth]{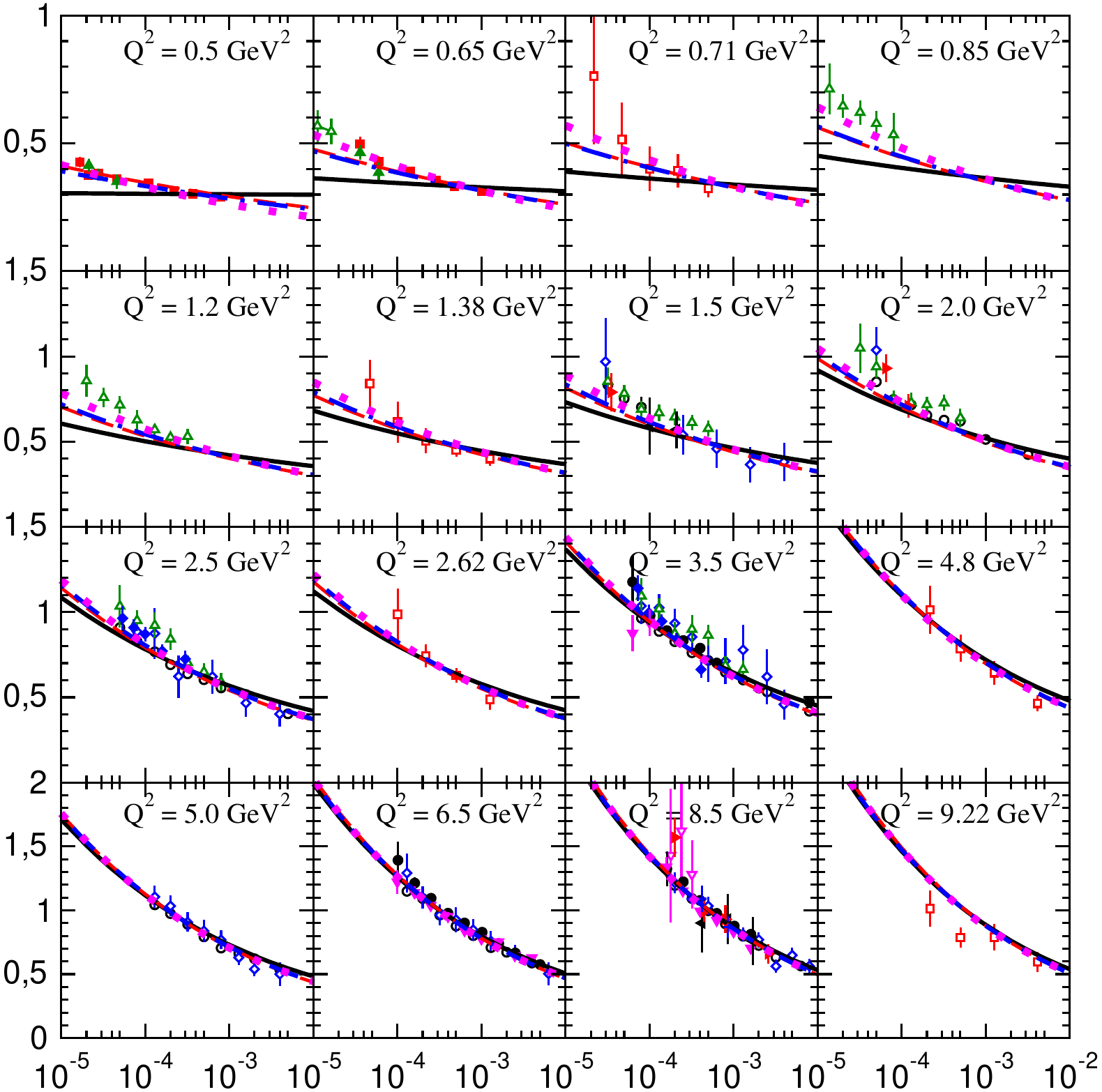}
\caption{$x$ dependence of $F_2(x,Q^2)$ in bins of $Q^2$.
The experimental data from H1 (open points) and ZEUS (solid points) are
compared with the NLO fits for $Q^2\geq0.5$~GeV$^2$ implemented with the
canonical (solid lines), frozen (dot-dashed lines), and analytic (dashed lines)
versions of the strong-coupling constant.
For comparison, also the results obtained in Ref.~\cite{Illarionov:2008}
through a fit based on the renormalon model of higher-twist terms are shown
(dotted lines).}
\label{fig:F2}
\end{figure}

\begin{figure}[t]
\includegraphics[width=\textwidth]{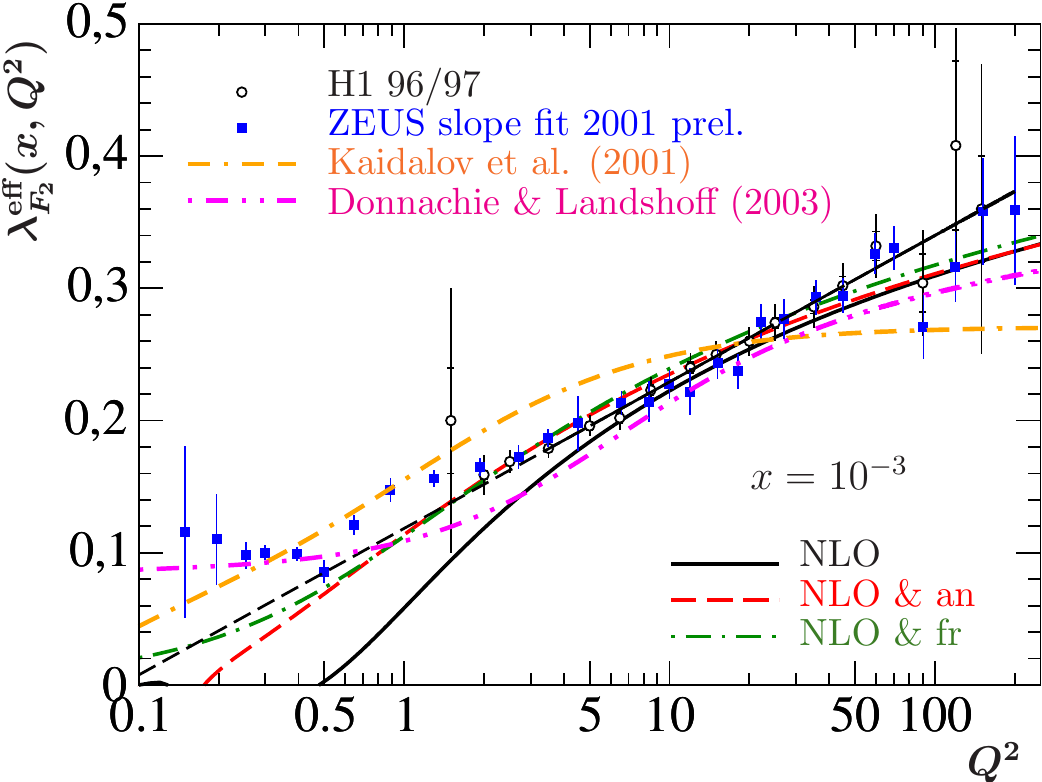}
\caption{$Q^2$ dependence of $\lambda^{\rm eff}_{F_2}(x,Q^2)$ for an average
small-$x$ value of $x=10^{-3}$.
The experimental data from H1 (open points) and ZEUS (solid points) are
compared with the NLO fits for $Q^2\geq0.5$~GeV$^2$ implemented with the
canonical (solid line), frozen (dot-dashed line), and analytic (dashed line)
versions of the strong-coupling constant.
The linear rise of $\lambda^{\rm eff}_{F_2}(x,Q^2)$ with $\ln Q^2$ as
described by Eq.~(\ref{1dd}) is indicated by the straight dashed line.
For comparison, also the results obtained in the phenomenological models by
Kaidalov et al.\ \cite{Kaidalov:2000} (dash-dash-dotted line) and by
Donnachie and Landshoff \cite{Donnachie:2003cs} (dot-dot-dashed line) are
shown.}
\label{fig:Q2-slope}
\end{figure}

\begin{figure}[t]
\includegraphics[width=\textwidth]{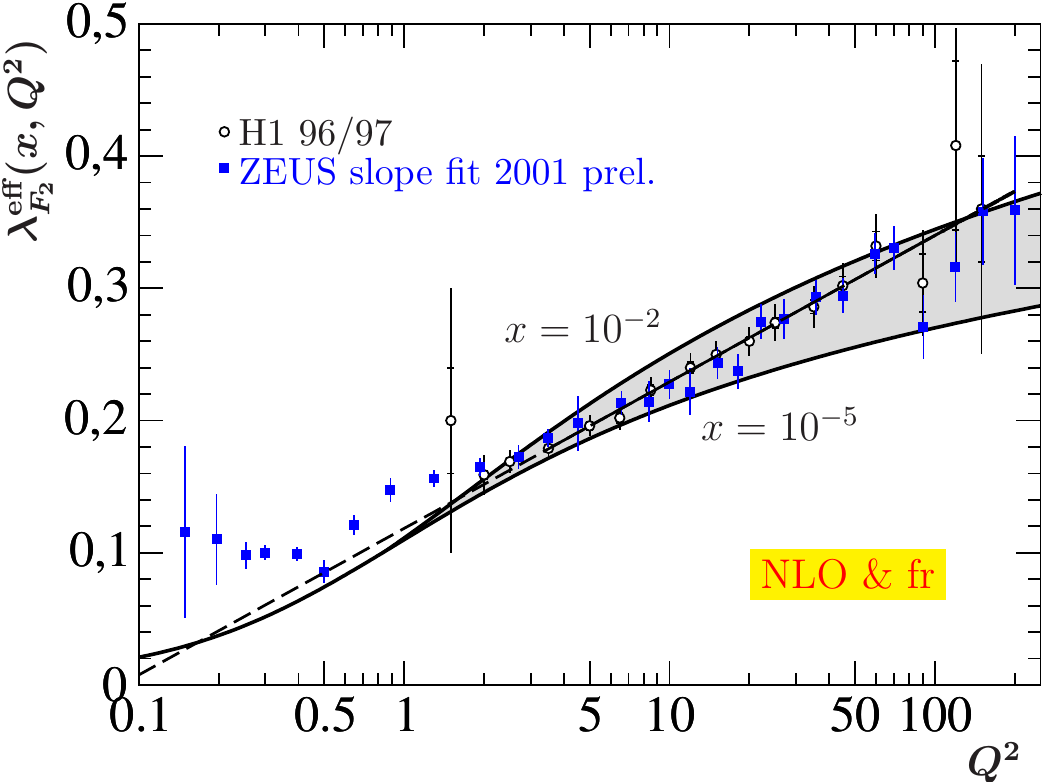}
\caption{Same as Fig.~\ref{fig:Q2-slope}, but for the NLO fit implemented with
the frozen version of the strong-coupling constant and for $x$ in the range
$10^{-5}<x<10^{-2}$.}
\label{fig:x-slope}
\end{figure}

Using the results of the previous sections, we analyze HERA data for the
structure function $F_2$ and its slope $\partial\ln F_2/\partial\ln (1/x)$
at small $x$ values from the H1 and ZEUS Collaborations
\cite{Abt:1993,Ahmed:1995,Aid:1996,Adloff:1997,Adloff:1999,Adloff:2001,%
Derrick:1993,Derrick:1995,Derrick:1996:C69,Derrick:1996:C72,Breitweg:1997,%
Breitweg:1999,Breitweg:2000,Chekanov:2001,%
Surrow:2002,Adloff:2001rw,Lastovicka:2002,Gayler:2002}
The experimental results for the $x$ dependence of $F_2$ in bins of $Q^2$ are
shown in Fig.~\ref{fig:F2}, while the $Q^2$ dependence of
$\lambda^{\rm eff}_{F_2}(x,Q^2)$ for an average small-$x$ value of $10^{-3}$
are shown in Figs.~\ref{fig:Q2-slope} and \ref{fig:x-slope}.

\begin{table}
\caption{\label{Tab:H1+ZEUS:96/97}\sffamily
Results of the LO and NLO fits to H1 and ZEUS data 
\cite{Abt:1993,Ahmed:1995,Aid:1996,Adloff:1997,Adloff:1999,Adloff:2001,%
Derrick:1993,Derrick:1995,Derrick:1996:C69,Derrick:1996:C72,Breitweg:1997,%
Breitweg:1999,Breitweg:2000,Chekanov:2001}
for different small-$Q^2$ cuts.}
\centering
\vspace{0.3cm}
\begin{tabular}{|l||c|c|c||r|} \hline \hline
& $A_g$ & $A_q$ & $Q_0^2$ [GeV$^2$] &
 $\chi^2 /{\rm n.d.f.}$ \\
\hline\hline
$Q^2 \geq 1.5$~GeV$^2$  &&&& \\
 LO & $0.784\pm0.016$ & $0.801\pm0.019$ & $0.304\pm0.003$ & 754/609 \\
 LO an. & $0.932\pm0.017$ & $0.707\pm0.020$ & $0.339\pm0.003$ & 632/609  \\
  LO fr. & $1.022\pm0.018$ & $0.650\pm0.020$ & $0.356\pm0.003$ & 547/609   \\
\hline
 NLO & $-0.200\pm0.011$ & $0.903\pm0.021$ & $0.495\pm0.006$ & 798/609 \\
 NLO an. & $0.310\pm0.013$ & $0.640\pm0.022$ & $0.702\pm0.008$ & 655/609  \\
  NLO fr. & $0.180\pm0.012$ & $0.780\pm0.022$ & $0.661\pm0.007$ & 669/609   \\
\hline\hline
 $Q^2 \geq 0.5$~GeV$^2$  &&&& \\
 LO & $0.641\pm0.010$ & $0.937\pm0.012$ & $0.295\pm0.003$ & 1090/662 \\
 LO an. & $0.846\pm0.010$ & $0.771\pm0.013$ & $0.328\pm0.003$ & 803/662  \\
  LO fr. & $1.127\pm0.011$ & $0.534\pm0.015$ & $0.358\pm0.003$ & 679/662   \\
\hline
 NLO & $-0.192\pm0.006$ & $1.087\pm0.012$ & $0.478\pm0.006$ & 1229/662 \\
 NLO an. & $0.281\pm0.008$ & $0.634\pm0.016$ & $0.680\pm0.007$ & 633/662  \\
  NLO fr. & $0.205\pm0.007$ & $0.650\pm0.016$ & $0.589\pm0.006$ & 670/662   \\
\hline \hline
\end{tabular}
\end{table}
In order to keep the analysis as simple as possible, we fix $f=4$ and
$\alpha^{\overline{\rm MS}}_s(M_Z^2)=0.1166$, so that
$\Lambda^{(4)}_{\overline{\rm MS}} = 284$~MeV and
$\Lambda^{(4)}_{\rm LO} = 112$~MeV, in agreement with the more recent ZEUS
results \cite{Chekanov:2001}.
We fit the combined H1 and ZEUS data on $F_2$
\cite{Abt:1993,Ahmed:1995,Aid:1996,Adloff:1997,Adloff:1999,Adloff:2001,%
Derrick:1993,Derrick:1995,Derrick:1996:C69,Derrick:1996:C72,Breitweg:1997,%
Breitweg:1999,Breitweg:2000,Chekanov:2001}
at LO and NLO imposing two different cuts on $Q^2$, namely $Q^2>1.5$~GeV$^2$
and $Q^2>0.5$~GeV$^2$.
The resulting values for $A_g$, $A_q$, and $Q_0^2$ are collected in
Table~\ref{Tab:H1+ZEUS:96/97} together with the values of $\chi^2$ per data
point ($\chi^2/{\rm n.d.f.}$) achieved.
In Fig.~\ref{fig:F2}, the H1 and ZEUS data on $F_2$, which come as $x$
distributions in bins of $Q^2$, are compared with the NLO result obtained with
the cut $Q^2>0.5$~GeV$^2$. 
Furthermore, the $Q^2$ dependence of $\lambda^{\rm eff}_{F_2}(x,Q^2)$ as
determined by H1 and ZEUS at an average small-$x$ value of $10^{-3}$ is
confronted with the result of the NLO fit for $Q^2>0.5$~GeV$^2$
in Fig.~\ref{fig:Q2-slope}.

Because the twist-two approximation is only reasonable at
$Q^2 \gsim 2.5$~GeV$^2$ \cite{Illarionov:2008}, as may be seen from
Fig.~\ref{fig:F2}, some theoretical improvements are necessary for smaller
$Q^2$ values.
In Ref.~\cite{Illarionov:2008}, the higher-twist corrections through twist six
were added to find good agreement for $Q^2 \gsim 0.5$~GeV$^2$.
However, the twist-four and twist-six terms increase the number of parameters,
which become strongly correlated.

Here, we investigate an alternative possibility, namely to modify the
strong-coupling constant in the infrared region.
Specifically, we consider two modifications, which effectively increase the
argument of the strong-coupling constant at small $Q^2$ values, in accordance
with
Refs.~\cite{Kotikov:1994:R,Shirkov:1995,Brodsky:1998,Ciafaloni:2000,%
Altarelli:2001,Andersson:2002}.
In the first case, which is more phenomenological, we introduce a freezing
of the strong-coupling constant by changing its argument as
$Q^2 \to Q^2 + M^2_{\rho}$, where $M_{\rho}$ is the rho-meson mass
\cite{Badelek:1996}.
Thus, in the formulae of Section~\ref{sec:DAS} 
and their NLO generalizations \cite{Kotikov:1999,Illarionov:2008},
we introduce the following replacement
\begin{equation}
\alpha^{\rm i}_s(Q^2) \to \alpha^{\rm i}_{\rm fr}(Q^2)
= \alpha^{i}_s(Q^2 + M^2_{\rho})
\qquad({\rm i}= {\rm LO}, \overline{\rm MS}),
\label{Intro:2}
\end{equation}
where $\alpha^{\rm LO}_s(Q^2)$ and $\alpha^{\overline{\rm MS}}_s(Q^2)$ have the
canonical forms dictated by the renormalization group.

The second possibility is based on the idea by Shirkov and Solovtsov 
\cite{Shirkov:1997,Solovtsov:1999} (see also the recent reviews in
Refs.~\cite{Bakulev:2008qq,Cvetic:2008,Stefanis:2009kv} and the references
cited therein) regarding the analyticity of the strong-coupling constant that
leads to an additional power dependence.
In this case, the one-loop and two-loop coupling constants
$\alpha^{\rm LO}_s(Q^2)$ and $\alpha^{\overline{\rm MS}}_s(Q^2)$ appearing
in the formulae of the previous sections and their NLO generalizations are to
be replaced as
\begin{align}
	\alpha^{\rm LO}_s(Q^2) &\to \alpha^{\rm LO}_{\rm an}(Q^2) =
		\alpha^{\rm LO}_s(Q^2) - \dfrac{1}{\beta_0}\,
\dfrac{\Lambda^2_{\rm LO}}{Q^2 - \Lambda^2_{\rm LO}},
\nonumber\\ 
	\alpha^{\overline{\rm MS}}_s(Q^2) &\to \alpha^{\overline{\rm MS}}_{\rm an}(Q^2) =
		\alpha^{\overline{\rm MS}}_s(Q^2) - \dfrac{1}{2\beta_0}\, 
\dfrac{\Lambda^2_{\overline{\rm MS}}}{Q^2 - \Lambda^2_{\overline{\rm MS}}} + \ldots  ,
\label{an:NLO}
\end{align}
where the ellipsis stands for cut terms which give negligible contributions.

We thus repeat the LO and NLO fits discussed above using in turn the frozen
and analytic versions of the strong-coupling constant according to the
replacements of Eqs.~(\ref{Intro:2}) and (\ref{an:NLO}), respectively.
The results for $F_2$ are included in Table~\ref{Tab:H1+ZEUS:96/97} and
Fig.~\ref{fig:F2} and those for $\lambda^{\rm eff}_{F_2}(x,Q^2)$ in
Fig.~\ref{fig:Q2-slope}.
Figure~\ref{fig:F2}, as also Fig.~5 in Ref.~\cite{Illarionov:2008}, only
covers the small-$Q^2$ region, $Q^2<9.22$~GeV$^2$, because the three NLO
predictions are hardly distinguishable at larger values of $Q^2$.

From Fig.~\ref{fig:F2}, we observe that the fits based on the frozen and
analytic strong-coupling constants are very similar and describe the data in
the small-$Q^2$ range significantly better than the canonical fit.
This is also reflected in the values of $\chi^2/{\rm n.d.f.}$ listed in
Table~\ref{Tab:H1+ZEUS:96/97}.
The improvement is especially striking at NLO if data with very small $Q^2$
values, with $Q^2 \geq 0.5$~GeV$^2$, are included in the fits.
Then ${\rm \chi^2 /{\rm n.d.f.}}$ is almost reduced by a factor of two to
assume values close to unity when the canonical version of the strong-coupling
constant is replaced by the frozen or analytic ones.
The situation is very similar to the case when the higher-twist corrections
according to the renormalon model are included \cite{Illarionov:2008}.
In order to illustrate this, we display in Fig.~\ref{fig:F2} also the results
obtained at NLO in the renormalon model of higher-twist terms, which are taken
from Fig.~5 in Ref.~\cite{Illarionov:2008}.
We see that the latter describe the experimental data slighly better for
0.65~GeV$^2\lsim Q^2\lsim2.0$~GeV$^2$ than the results obtained here, which is
also reflected in the values of $\chi^2 /{\rm n.d.f.}$ achieved, namely
$\chi^2 /{\rm n.d.f.}=565/658=0.86$ for renormalon improvement versus
$\chi^2 /{\rm n.d.f.}=633/662=0.96$ and $670/662=1.01$ for analytic and frozen
strong-coupling constants, respectively.
However, one should bear in mind that this improvement happens at the expense
of introducing four additional adjustable parameters.

Figure~\ref{fig:Q2-slope} nicely demonstrates that the theoretical description
of the small-$Q^2$ ZEUS data on $\lambda^{\rm eff}_{F_2}(x,Q^2)$ by NLO QCD is
significantly improved by implementing the frozen and analytic strong-coupling
constants.
Again, these two alternatives lead to very similar results.
For comparison, the linear rise of $\lambda^{\rm eff}_{F_2}(x,Q^2)$ with
$\ln Q^2$ as described by Eq.~(\ref{1dd}) is also indicated in 
Fig.~\ref{fig:Q2-slope}.
For comparison, we display in Fig.~\ref{fig:Q2-slope} also the results
obtained by Kaidalov et al.\ \cite{Kaidalov:2000} and by Donnachie and
Landshoff \cite{Donnachie:2003cs} adopting phenomenological models based on
Regge theory.
While they yield an improved description of the experimental data for
$Q^2\lsim0.4$~GeV$^2$, the agreement generally worsens in the range
2~GeV$^2\lsim Q^2\lsim8$~GeV$^2$.

As may be seen from Table~\ref{Tab:H1+ZEUS:96/97}, the three NLO fits for
$F_2(x,Q^2)$ yield $Q^2_0 \approx 0.5$--$0.7$~GeV$^2$ (see also
Ref.~\cite{Illarionov:2008}).
Figure~\ref{fig:Q2-slope} shows that the conventional NLO fit yields
$\lambda^{\rm eff}_{F_2}(x,Q^2_0)=0$ as suggested by Eq.~(\ref{1}).
The replacements of Eqs.~(\ref{Intro:2}) and (\ref{an:NLO}) raise the value
of $\lambda^{\rm eff}_{F_2}(x,Q^2_0)$.
In fact, the results for $\lambda^{\rm eff}_{F_2}(x,Q^2)$ obtained with the
frozen and analytic versions of the strong-coupling constant agree much better
with the ZEUS data at $Q^2\gsim 0.5$~GeV$^2$.
Nevertheless, for $Q^2 < 0.5$~GeV$^2$, there is still some disagreement with
the data, which needs additional investigation.

In Fig.~\ref{fig:Q2-slope}, the NLO results for
$\lambda^{\rm eff}_{F_2}(x,Q^2)$ are evaluated at $x=10^{-3}$.
In Fig.~\ref{fig:x-slope}, we study the variation with $x$ in the range
$10^{-5}<x< 10^{-2}$.
For simplicity, we only do this for the case of the frozen strong-coupling
constant; the result for the analytic one would be very similar.
We observe good agreement between the experimental data and the generalized
DAS approach for a broad range of small-$x$ values.
At small $Q^2$ values, $\lambda^{\rm eff}_{F_2}(x,Q^2)$ is practically
independent of $x$, which is because the variable $\rho$ defined in
Eq.~(\ref{slo}) takes rather small values there.
At large $Q^2$ values, the $x$ dependence of $\lambda^{\rm eff}_{F_2}(x,Q^2)$
is rather strong.
However, it is well known that the boundaries and mean values of the
experimental $x$ ranges \cite{Adloff:2001rw} increase proportionally with
$Q^2$, which is related to the kinematical restrictions in the HERA
experiments, namely $x \sim 10^{-4} \times  Q^2$
(see Refs.~\cite{Adloff:1997,Adloff:2001,Breitweg:2000,Chekanov:2001} and, for
example, Fig.~1 of Ref.~\cite{Surrow:2002}).
From Fig.~\ref{fig:x-slope}, we see that the HERA data are close to
$\lambda^{\rm eff}_{F_2}(x,Q^2)$ at $x \sim 10^{-4}$--$10^{-5}$
for $Q^2=4$~GeV$^2$ and at $x \sim 10^{-2}$ for $Q^2=100$~GeV$^2$.
Indeed, the correlations between $x$ and $Q^2$ of the form 
$x_{\rm eff}= a \times 10^{-4} \times Q^2$ with $a=0.1$ and $1$ lead to a
modification of the $Q^2$ evolution which starts to resemble $\ln Q^2$, rather
than $\ln \ln Q^2$ as is standard \cite{Kotikov:1997:JETP}.

\section{Conclusions}
\label{sec:concl}

We studied the $Q^2$ dependence of the structure function $F_2$ and the slope 
$\lambda^{\rm eff}_{F_2}=\partial\ln F_2/\partial\ln (1/x)$ at small $x$
values in the framework of perturbative QCD.
Our twist-two results are in very good agreement with HERA data 
\cite{Abt:1993,Ahmed:1995,Aid:1996,Adloff:1997,Adloff:1999,Adloff:2001,%
Derrick:1993,Derrick:1995,Derrick:1996:C69,Derrick:1996:C72,Breitweg:1997,%
Breitweg:1999,Breitweg:2000,Chekanov:2001,%
Surrow:2002,Adloff:2001rw,Lastovicka:2002,Gayler:2002}
at $Q^2 \gsim 2.5$~GeV$^2$, where perturbation theory is applicable.
The applications of the frozen and analytic versions of the strong-coupling
constants, $\alpha^{\overline{\rm MS}}_{\rm fr}(Q^2)$ and
$\alpha^{\overline{\rm MS}}_{\rm an}(Q^2)$, significantly improve the agreement
with the HERA data
\cite{Abt:1993,Ahmed:1995,Aid:1996,Adloff:1997,Adloff:1999,Adloff:2001,%
Derrick:1993,Derrick:1995,Derrick:1996:C69,Derrick:1996:C72,Breitweg:1997,%
Breitweg:1999,Breitweg:2000,Chekanov:2001,%
Surrow:2002,Adloff:2001rw,Lastovicka:2002,Gayler:2002}
for both the structure function $F_2$ and its slope
$\lambda^{\rm eff}_{F_2}(x,Q^2)$ for small $Q^2$ values,
$Q^2 \gsim 0.5$~GeV$^2$.
The results obtained with these infrared-modified strong-coupling constants
and also those based on the renormalon model with higher-twist terms
incorporated, which were considered in Ref.~\cite{Illarionov:2008}, are very
similar numerically.

As a next step of our investigations, we plan to fit the HERA data 
\cite{Abt:1993,Ahmed:1995,Aid:1996,Adloff:1997,Adloff:1999,Adloff:2001,%
Derrick:1993,Derrick:1995,Derrick:1996:C69,Derrick:1996:C72,Breitweg:1997,%
Breitweg:1999,Breitweg:2000,Chekanov:2001}
for $F_2(x,Q^2)$ using alternative analytic versions of the strong-coupling
constant (see, for example, the recent reviews in
Refs.~\cite{Bakulev:2008qq,Cvetic:2008,Stefanis:2009kv}),
to find out if the theoretical description of the slope 
$\lambda^{\rm eff}_{F_2}$ can be further improved at small $Q^2$ values.

\section*{Acknowledgments}

The work of G.C. was supported in part by Chilean Fondecyt Grant No.~1095196.
The work of B.A.K. was supported in part by the German Federal Ministry for
Education and Research BMBF through Grant No.\ 05~HT6GUA and by the Helmholtz
Association HGF through Grant No.\ HA~101.
The work of A.V.K. was supported in part by the German Research Foundation DFG
through Grant No.\ INST 152/465--1, 
by the Heiserberg-Landau Program through Grant No.~5,
and by the Russian Foundation for Basic Research through Grant
No.~08--02--00896--a.

\bibliographystyle{revtex}
\bibliography{hep}

\end{document}